\documentclass[letterpaper]{article} 
\usepackage{aaai24}  
\usepackage{times}  
\usepackage{helvet}  
\usepackage{courier}  
\usepackage[hyphens]{url}  
\usepackage{graphicx} 
\usepackage{natbib}  
\usepackage{caption} 

\usepackage{algorithm}
\usepackage{algorithmic}
\usepackage[aboveskip=1pt,labelfont=bf,labelsep=period,justification=raggedright,singlelinecheck=off]{caption}

\newenvironment{quoteitalicized}
    {\begin{quote}}
    {\end{quote}}
\newcommand{\quotes}[2]{\begin{quoteitalicized}\small \textit{#1} {#2}\end{quoteitalicized}}
\usepackage{makecell}

\usepackage{newfloat}
\usepackage{listings}
\DeclareCaptionStyle{ruled}{labelfont=normalfont,labelsep=colon,strut=off} 
\floatstyle{ruled}
\newfloat{listing}{tb}{lst}{}
\floatname{listing}{Listing}
\usepackage{dirtytalk}
\usepackage{xspace}
\usepackage{booktabs}
\usepackage{cleveref}
\usepackage{multirow}
\usepackage[inline]{enumitem}
\usepackage{xcolor}
\usepackage{soul}

\title{Social Scientists on the Role of AI in Research}
\author{
    Tatiana Chakravorti\textsuperscript{\rm 1},
    Xinyu Wang\textsuperscript{\rm 1},
    Pranav Narayanan Venkit\textsuperscript{\rm 1},
    Sai Koneru\textsuperscript{\rm 1},
    Kevin Munger\textsuperscript{\rm 2},
    Sarah Rajtmajer\textsuperscript{\rm 1}
}
\affiliations{
    \textsuperscript{\rm 1}Pennsylvania State University, USA\\
    \textsuperscript{\rm 2}European University Institute, Florence\\

}

\usepackage{bibentry}

\begin{document}

\maketitle

\begin{abstract}
The integration of artificial intelligence (AI) into social science research practices raises significant technological, methodological, and ethical issues. We present a community-centric study drawing on 284 survey responses and 15 semi-structured interviews with social scientists, describing their familiarity with, perceptions of the usefulness of, and ethical concerns about the use of AI in their field. A crucial innovation in study design is to split our survey sample in half, providing the same questions to each -- but randomizing whether participants were asked about "AI" or "Machine Learning" (ML). We find that the use of AI in research settings has increased significantly among social scientists in step with the widespread popularity of generative AI (genAI). These tools have been used for a range of tasks, from summarizing literature reviews to drafting research papers. Some respondents used these tools out of curiosity but were dissatisfied with the results, while others have now integrated them into their typical workflows. Participants, however, also reported concerns with the use of AI in research contexts. This is a departure from more traditional ML algorithms which they view as statistically grounded. Participants express greater trust in ML, citing its relative transparency compared to black-box genAI systems. Ethical concerns, particularly around automation bias, deskilling, research misconduct, complex interpretability, and representational harm, are raised in relation to genAI. We situate these findings within broader sociotechnical debates, arguing that responsible integration of AI in social science research requires more than technical solutions: it demands a rethinking of research values, human-centered design, and institutional support structures. To guide this transition, we offer recommendations for AI developers, researchers, educators, and policymakers focusing on explainability, transparency, ethical safeguards, sustainability, and the integration of lived experiences into AI design and evaluation processes. 
  
\end{abstract}

\section{Introduction}



The rapid advancement of artificial intelligence (AI) and machine learning (ML) technologies is transforming the landscape of social science scholarship and research \cite{bail2024can}. The emergence of large language models (LLMs) such as ChatGPT and Gemini 
promise new efficiencies but bring challenges as well \cite{salganik2019bit, ziems2024can}. The education sector and academia have felt the immediate impact of LLMs on teaching practices, experimental workflows, manuscript writing, and editorial processes \cite{celik2022promises, wu2024reacting,omran2024redefining}. 
While, researchers have increasingly adopted these tools for a wide range of applications, including data collection, annotation, and analyses, as well as a variety of coding tasks previously inaccessible to researchers without extensive programming backgrounds 
\cite{zhang2023generative, dandamudi2025advancing,sadiku2021artificial, fan2023interpretable, deranty2024artificial, obreja2024mapping, bail2024can}. Researchers are exploring entirely new paradigms, such as simulating social interactions, generating synthetic interview data, and prototyping new theories \cite{grossmann2023ai}. 

However, the adoption of AI into the social sciences is not a simple upgrade, but a fundamental sociotechnical shift, entailing negotiations across practical, methodological, and ethical lines. AI and ML have the potential to produce or reinforce existing biases \cite{o2017weapons, bender2021dangers}. Design choices shape what kinds of knowledge are produced, whose perspectives are prioritized, and how research problems are framed \cite{selbst2019fairness}. 
Further, the social sciences are methodologically diverse, including qualitative, quantitative, and mixed methods, e.g., ethnography, interviews, experiments, agent-based modeling, and observational studies \cite{wulff2022bridging, ligo2021comparing}. This diversity presents unique challenges and opportunities when integrating AI, into the epistemological standards and interpretive needs of different sub-fields. 

Our work seeks to understand how social science researchers are navigating AI and ML, scaffolded by 
the following research questions (RQs):

    \begin{itemize}
    \item\textbf{RQ1}: How are AI and ML reshaping research practices in the social sciences?
    \item\textbf{RQ2}: What are the common ethical, methodological, and practical concerns among social scientists regarding the use of AI and ML in research?
    \item\textbf{RQ3}: How are social scientists negotiating the role of human judgment in the context of increasingly autonomous tools?
    \end{itemize}

Drawing on a \textit{mixed-methods} approach that includes surveys followed by semi-structured interviews, this study explores how social science researchers navigate excitement and tensions around automation when working with AI and ML. The same survey was given to all participants, \textit{with one crucial element randomized: half of the participants were asked their perceptions about "AI," and the other half about "ML." This design allows us to differentiate these two often-conflated developments in social science practice.}

We find that the use of AI in research settings is common among social scientists, particularly with the growing popularity of genAI. These tools are employed for a variety of tasks, from summarizing literature to drafting research papers, whereas ML was used for mostly prediction or classification tasks. Participants raised concerns about de-skilling, lack of standardization, ethical risks, bias, model interpretability, inadequate training, and over-reliance on AI systems. \textit{Ethical concerns were less pronounced when we asked about ML techniques than when we asked about AI -- suggesting evolution in the distribution of attention with the rise of genAI.} These insights underscore the importance of interdisciplinary collaboration and community-centered research paradigms, where AI integration is informed by the experiences and practical needs of social scientists. To support this shift, we propose actionable recommendations for four key stakeholder groups: developers; researchers; educators; and policymakers.
Ultimately, we advocate for reimagining AI not as a replacement for human insight, but as a collaborator in processes of knowledge production.

\section{Related Work}
\subsection{AI and ML in the social sciences}
Artificial intelligence and machine learning are actively transforming scientific processes across the research lifecycle \cite{chubb2022speeding, xu2021artificial}. 
AI integration into the social sciences has facilitated analyses of large scale datasets, uncovering novel insights and opening new research directions \cite{grimmer2015we, grossmann2023ai}. This integration has been accompanied by robust discussions about the significance of the moment and questions about the role these technologies should play in social science research  
\cite{iorga2023sociological, williams2024paradigm}.

Initially, ML was viewed as a powerful collection of computational tools required for navigating the rapidly expanding availability of ``big data" \cite{salganik2019bit}. Given that new datasets pose challenges for conventional statistical analyses due to scale or complexity, ML techniques were framed primarily as advanced methods for pattern discovery, prediction, and classification \cite{varian2014big, grimmer2021machine}. This positioned ML as an extension of the quantitative toolkit, a required adaptation to deal with new types of evidence \cite{einav2014economics}. 

As integration grew, the role of ML began to diversify and sometimes generated discussions centering on the distinction between the predictive strength of ML models and the traditional emphasis of social sciences research on causal explanation \cite{hofman2021integrating,shmueli2010explain}. Reflecting these discussions, efforts emerged to adapt ML techniques specifically for causal analysis, seeking ways to improve the estimation and understanding of causal relationships, particularly when working with complex observational data \cite{athey2017state, chernozhukov2018double}. Alongside these developments, the language used to describe these approaches often remained fluid and overlapping. Researchers might find the distinctions between ``ML", ``AI," ``computational social science," and advanced statistics fuzzy, with usage varying depending on their particular training or the specific domain \cite{edelmann2020computational, grimmer2021machine}.

The advent of generative AI, particularly large language models, has arguably triggered a significant shift in interpretation \cite{zhang2023generative}. For many researchers, "AI" is now increasingly understood through the lens of these generative capabilities – systems that can produce human-like text, code, and interact in complex ways \cite{bail2023can}. This interpretation casts AI not merely as an analytical tool operating on existing data, but as an active generator of content and potentially even a collaborator or simulator of human behavior \cite{anthis2025llm, jansen2023employing, dominguez2024questioning}



\subsection{Ethical concerns about AI}

As AI and ML technologies become increasingly embedded in research and daily life, conversations about their benefits are frequently accompanied by growing concerns about their potential to cause harm.  One of the most pressing and recurring concerns is bias—both in how these systems are developed and how they are deployed \cite{o2017weapons, bender2021dangers}. These issues are exacerbated by the opaque, `black-box' nature of many AI systems, which are often proprietary and lack transparency in terms of architecture, training methodology, and data provenance \cite{o2017weapons, bender2018data, castelvecchi2016can, von2021transparency}. At the heart of these concerns lies the data that is used to train and develop these systems \cite{hacker2021legal, sambasivan2021everyone}. GenAI models, especially large language models, are typically trained on vast corpora scraped from the internet and other publicly available sources \cite{narayanan2023unmasking, paullada2021data}. While this scale and diversity might suggest objectivity or comprehensiveness, these datasets tend to amplify dominant narratives—often Western, English-speaking, and majority-population-centric—while marginalizing or misrepresenting voices from minoritized communities \cite{gautam2024melting, das2024colonial}. This results in representational harms that can take the form of stereotyping, cultural misappropriation, and the erasure of difference \cite{dev2022measures, blodgett2022responsible}. Bias in AI is not limited to model outputs but permeates every layer of these systems, from embedding representations to classification algorithms to generative models \cite{bolukbasi2016man, mehrabi2021survey, ntoutsi2020bias}. 

\subsection{Sociotechnical concerns about AI}
As AI systems become more integrated into decision-making pipelines, the illusion that they are solely  technical systems becomes harder to sustain. Researchers have noted how technical design decisions intersect with social contexts, reinforcing existing inequalities or introducing new forms of algorithmic harm from biased behaviors \cite{dolata2022sociotechnical, narayanan2023towards, ehsan2021expanding}. For instance, in tasks like recidivism prediction, toxicity detection, and facial recognition, biased outputs can lead to discriminatory outcomes when these systems are deployed in real-world social infrastructures \cite{bender2021dangers, o2017weapons}. 
Work in HCI and science and technology studies (STS) examines these harms through the lens of sociotechnical entanglement, emphasizing how users, designers, and affected communities experience and respond to algorithmic systems. Research frameworks now increasingly move beyond surface-level performance metrics to interrogate how harm materializes through misinformation, stereotyping, and exoticism—phenomena that not only distort representation but also shape how entire communities are understood, regulated, or excluded \cite{ghosh2024generative, dev2022measures, blodgett2022responsible}.

\textit{Much of the existing literature has examined harms at the societal or individual user level, yet there remains a gap in understanding how AI technologies impact academic efforts and usage, particularly within the social sciences.} These disciplines, which traditionally rely on interpretive, qualitative, and mixed-method approaches, are being reshaped by tools that offer automation, scale, and new modalities of expression. We respond by offering a detailed, mixed-methods investigation into how social science researchers perceive, adopt, and contend with AI and ML technologies. We explore how scholars navigate the affordances and the ethical entanglements of these tools, focusing on impacts across research design, authorship, epistemology, and judgment. 

\section{Methods}

\subsection{Study design}
We adopted an \textit{explanatory sequential mixed method technique} \cite{ivankova2006using} wherein quantitative (survey) data is collected and analyzed first, followed by qualitative data collection to explore these findings in further depth. 
 \emph{The complete survey instruments (both versions) and the interview topic guide are provided in Supplemental Materials.}
 
\noindent \textbf{Survey design.} Our survey followed standard exploratory survey design methodology \cite{rosen2013media, baker2016reproducibility, van2023ai, chakravorti2025reproducibility}, with one crucial innovation. \textit{The survey questionnaire was held constant, but the target of inquiry was randomized; one-half of the surveys asked about ``Artificial Intelligence," and the other half of the surveys asked all the same questions about ``Machine Learning" instead of AI}. Our aim was to understand how social science researchers' perceptions, acceptance, and ethical considerations changed over the significant advancement in ML to generative AI. 

The survey was organized as follows. Participants were presented with information about the project and consented to participate. \textbf{Demographic information:} Participants were asked to provide information related to gender, age, ethnicity, academic position, field of research, and years of experience. The body of the survey contained 4 closed-ended questions and 5 open-ended questions. \textbf{Closed-ended questions:} These asked participants' to rate their usage of, acceptance of, and familiarity with AI/ML. \textbf{Open-ended questions:} These aimed to complement quantitative data with stated perceptions, experiences, and ethical and methodological challenges using AI/ML technologies in their research. A pilot version of the survey (both variants) was created and pretested by 5 social science researchers before deployment.

\noindent \textbf{Interview design.} All interviews were carried out via Zoom. The length of interviews ranged from 15 to 40 minutes; the majority were completed within 30 minutes.

Interviews were organized as follows. \textbf{Understanding AI and ML:} This set of questions is asked to understand what are AI and ML in terms of their understanding and how they differentiate these two from each other as we have not asked for any definition of them during the survey. \noindent \textbf{Ethical concerns about AI and ML:} With these questions we investigated is the ethical concerns and trust really increase with AI with elaborated examples to validate our survey findings. \noindent \textbf{Value of human judgment:} In this section, we asked participants how they are currently using these techniques and how they perceive the role of human judgment in an increasingly autonomous world. We also explored their preferences regarding the types of systems they choose to use. 

\subsection{Participant recruitment and compensation}
\noindent \textbf{Survey recruitment.} We recruited research-active academics across rank, i.e., assistant, associate, and full professors, post-doctorates and research associates, and graduate students working in US universities in the social sciences. We selected 30 universities randomly from the top 100 based on the US News and World Report (2003) rankings. 

We directly emailed researchers listed on departmental web pages, targeting the following disciplines: economics; political science; education; psychology; sociology; and marketing. We collected email IDs via web scraping of Universities' directories. \textit{In total, we emailed 8,000 social science researchers. One-half received the survey on AI, while the other half received a survey on ML. We received 348 responses}; however, some of these responses were incomplete and thus excluded from our analysis. We only considered responses with a 100\% completion rate for the closed-ended questions, resulting in \textbf{284 valid responses}, or 3.56\% of the total surveys sent (143 AI survey; 141 ML survey). 

At the end of the survey, participants were given the opportunity to be entered into a draw to win one of ten (10) \$100 gift vouchers. 

\noindent \textbf{Interview recruitment.} Unrelated to the compensation opportunity, survey participants were invited to provide their email IDs for a follow-up interview study.  
We emailed all the participants with a stated interest in the interview study and conducted \textbf{15 semi-structured interviews} \cite{varanasi2022feeling, thakkar2022machine} to further validate quantitative findings from the survey study. Each participant was compensated \$30 for their time.

\subsection{Participant demographics}
Demographic characteristics of survey participants is presented in Table 1. 
Responses were received from individuals across disciplines and academic ranks. Our respondents are not representative of the US population; however, there are no significant differences in demographics across the ML and AI groups. Only one person from the ML survey did not disclose the demographic details but he did provide other responses. Demographic data for the 15 interview participants is provided in Table 3.

\subsection{Data analysis}
For the \textit{quantitative data analysis}, closed-ended questions were analyzed using \textbf{descriptive statistics}\cite{kaur2018descriptive} and exploratory data visualizations. For the \textit{qualitative data analysis}, first two authors analyzed the semi-structured interview transcripts using \textbf{thematic analysis} as outlined by Blandford \cite{blandford2016qualitative}. This qualitative data analysis method involves thoroughly reading the transcripts to identify patterns across the dataset and derive themes related to the research questions. We employed a collaborative and iterative coding process \cite{ding2022uploaders, huang2020you}. Initially, the first two authors read the interview transcripts multiple times to become familiar with the data. Open coding was then conducted to identify initial codes by the first and second authors. These codes were then organized into categories based on similarities and relationships. Finally, we refined these categories, assigned names to each theme, and developed a conceptual framework to address our research questions with input from the entire team. A similar approach was taken for the open-ended survey responses. 

\section{RQ1: AI and ML reshaping social science research practices}
\subsection{Understanding of AI and ML}
In the survey, participants were asked about their familiarity with AI and ML. Figure 1 shows the differences across our four outcome variables. Among respondents to the ML survey, 41.84\% reported being slightly familiar, and 7.09\% reported not being familiar at all with ML techniques. In contrast, familiarity was higher in the AI survey: only 22.37\% of respondents indicated slight familiarity and just 3.49\% reported no familiarity with AI techniques. We also examined how frequently respondents use AI/ML in their research. Notably, 40.43\% had never used ML techniques, compared to 27.27\% who had never used AI techniques.

\begin{table}[ht]
\small
\caption{Survey respondent demographics by survey type (ML: $n=141$, AI: $n=143$)}
\centering
\begin{tabular}{p{1.8cm}|p{2.5cm}|p{1.2cm}|p{1.2cm}}
\toprule
\textbf{\makecell{Characteristic}} & \textbf{\makecell{Category}} & \textbf{\makecell{ML \\Survey($n$)}} & \textbf{\makecell{AI \\Survey($n$)}} \\ \midrule
\multirow{5}{*}{\textbf{Gender}} & Female & 71 & 74 \\
& Male & 62 & 63 \\
& Non-binary & 4 & 3 \\
& Prefer not to say & 3 & 3 \\
& Others & 1 & 0  \\ 
\midrule
\multirow{4}{*}{\textbf{Ethnicity}} & White & 88  & 98  \\
& Asian & 23 & 24  \\
& Others & 20 & 12 \\
& Black or African American & 10 & 9 \\ \midrule
\multirow{6}{*}{\textbf{\makecell{Academic\\ Rank}}} & Graduate Student & 54 & 65 \\
& Assistant Professor & 18  & 14  \\
& Associate Professor & 19  & 22  \\
& Professor & 35  & 34  \\
& Others & 15  & 7  \\
& Postdoc & 0 & 1  \\ 
\midrule
\multirow{4}{*}{\textbf{Age Group}} & $<$25 & 10  & 14  \\
& 25--35 & 52  & 56  \\
& 36--45 & 30  & 28  \\
& $>$45 & 49  & 45  \\
\bottomrule
\end{tabular}
\label{tab:demo}
\end{table}


Fig \ref{fig:regression} displays the results of eight separate regressions, two (one about AI, one about ML) for each of the four outcome variables. The most statistically significant finding emerges in the position-related predictors, each of which is displayed relative to PhD students (the reference category). Here, \textit{there is no significant variation in attitudes towards ML across career stages -- but there is significant variation in attitudes towards AI. PhD students report greater acceptance (bottom left panel) than social scientists at any other career stage. They also report a higher frequency of usage and perceive AI to be more useful than either assistant or full professors.}

Gender dynamics present another intriguing dimension. \textit{Female respondents report lower frequency using and lower familiarity with both AI and ML, compared to male respondents.} There are not, however, any differences when we ask about acceptance or perceived usefulness of these technologies. The non-binary or alternative gender category shows additional variation, though the confidence intervals suggest less certainty about the precise nature of this relationship.
There are no significant differences between white and non-white respondents; we lack the sample size to break the latter category down into more granular detail.

\begin{figure}[h]
  \centering 
   \includegraphics[width=.99\linewidth]{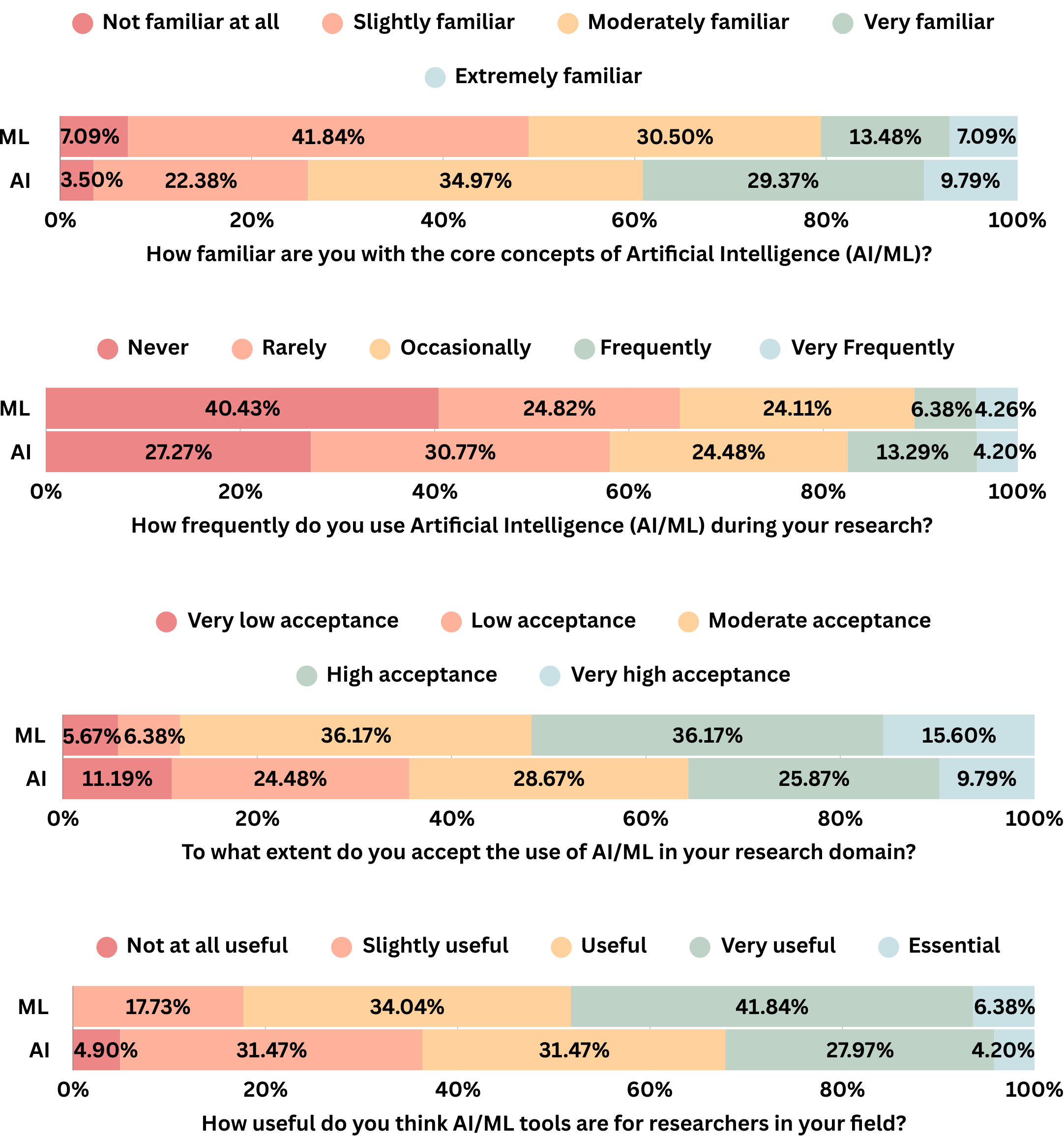} 
   \small
   \caption{Familiarity, frequency of use, acceptance, usefulness of AI/ML technologies}
   \label{fig:crosstab}
 \end{figure}

\begin{figure}[h]
  \centering 
   \includegraphics[width=0.99\linewidth]{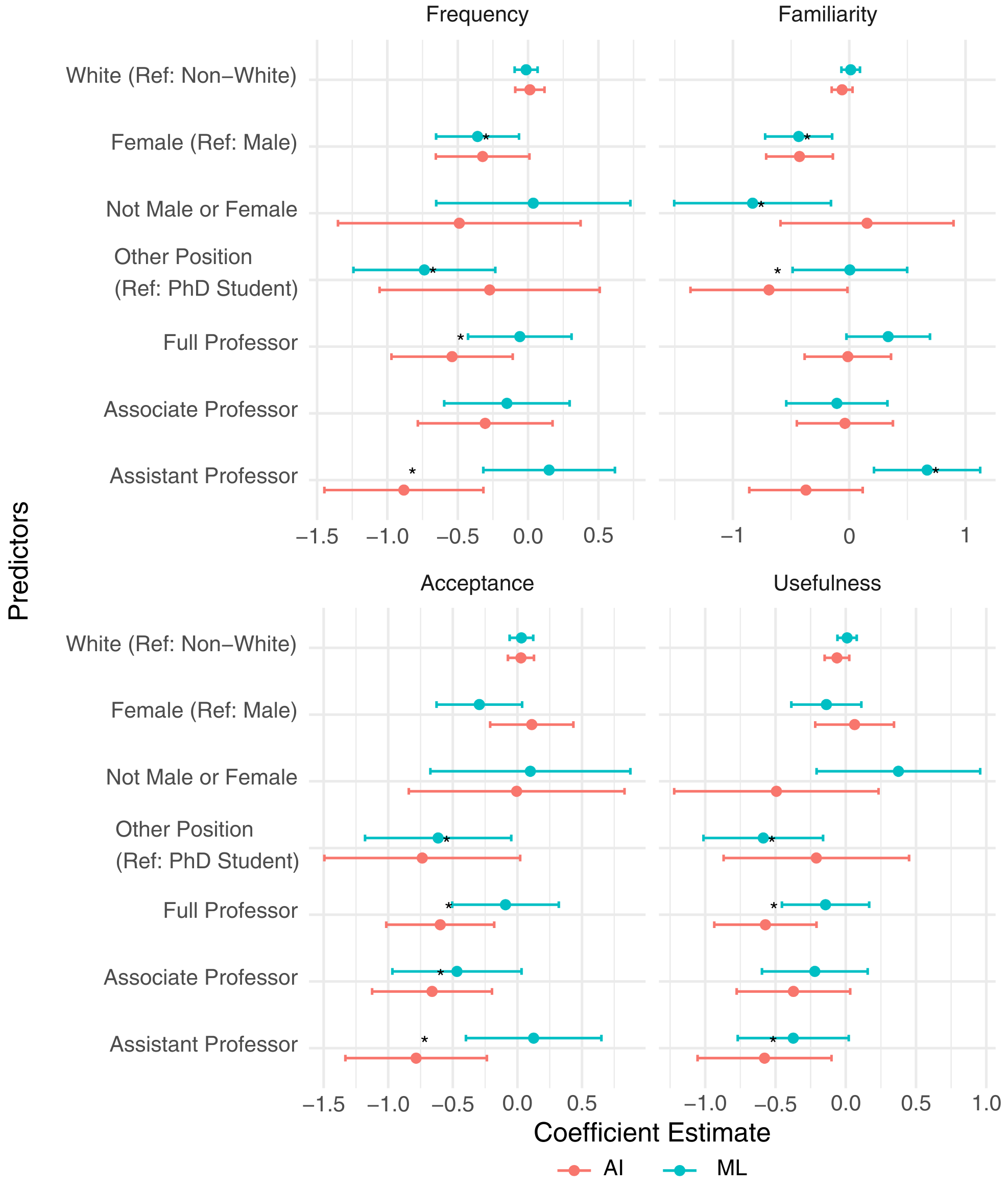} 
   \small
   \caption{Regression analysis based on demographics of the frequency, familiarity, acceptance, and usefulness of AI and ML methods}
   \label{fig:regression}
 \end{figure}

In the follow-up interviews, we explored how social science researchers perceive AI and ML technologies---and what those terms mean to them. 
We observed that when participants were asked to explain AI in their own words, they primarily thought of generative AI tools or large language models. They tended to view machine learning as an algorithm or method that helps with classification and prediction for different applications, while they imagined AI as a product or as a tool that uses large language models and can mimic human behavior. 

\quotes{I think that when we are talking about AI, in the recent discourse, we are talking about the large language models. I use a lot of ChatGPT in particular for all sorts of possible prompts.} {--P1}


P10 characterized AI deep-learning algorithms leveraging large datasets. P10 had a much deeper understanding of AI and described LLMs as a type of AI working with language. Except for P3, P5, and P7, other participants reported that LLMs come to their minds first when they hear AI. Participant P5 differed from others; as she sees AI as ChatGPT and doesn't have any understanding of ML.  

\quotes{It is typically, electronically, some type of system like ChatGPT or another brand to ask questions, and prompts, and have answers spit out at you. I have no idea what is machine learning is.} {--P5}

All participants except P5 mentioned ML is a part of AI but all AI methods do not belong to ML. They all mentioned AI has much broader and more advanced applications. Some participants have hesitations about their understanding: 

\quotes{I don't know if I know the nuance difference between artificial intelligence and machine learning. I think of machine learning as being much more predictive or providing output based on a studied corpus of text or a body. I assume artificial intelligence probably goes beyond that in scope. But I'm not sure I know the nuanced difference.}{--P7}

\subsection{Usefulness and acceptance of AI and ML}
When participants were asked about their acceptance of AI/ML technologies in their research domain during the survey, we found that ML was better accepted than AI. A combined 51.77\% of respondents expressed high or very high acceptance of ML, compared to 35.66\% for AI. Similarly, perceptions of usefulness showed that 41.84\% of respondents considered ML to be very useful, while only 27.97\% said the same for AI. This is described in Figure 1. 
\textit{These differences were statistically significant (p $<$ .01). All statistical results are described in Table 2.}

Survey participants also shared how they use AI tools to support specific research tasks, particularly those related to information processing and literature review. Several responses pointed to the usefulness of AI tools for summarizing research papers:

\quotes{I have used Consensus to initiate a brief (informal) search — I appreciate being able to quickly pull several relevant papers (with summarization as part of the tool) and then use those as a starting point for a more manual search.}{--AI Survey}


Others described using AI tools to explore unfamiliar domains, identify key scholars, or clarify concepts:

\quotes{Yes, I have used ChatGPT to identify top scholars and seminal works in literature I am not familiar with.}{--AI Survey}

All interview participants noted that the emergence of generative AI tools has significantly increased the use of AI in social science research---from summarization and literature review, to writing codes, and drafting the text of research papers. Participant P1 uses genAI tools for coding and he reported it helps a lot with debugging. P10 also mentioned that he used ChatGPT specifically to debug code in R. He also mentioned that occasionally he used it for idea generation. 

\quotes{I try to use ChatGPT pretty conservatively… mostly for helping me debug code or to generate an idea.}{--P10}

P9 sees potential in genAI tools for tasks like transcription or first-pass annotation. But she also mentioned that she does not think these tools are good for writing papers or conducting qualitative analysis, and emphasizes that it is not yet expected or normalized in her field. On the other hand, P8 mentioned that he uses genAI tools to help automate literature review tasks, like extracting study objectives from papers, after validating a subset of results manually. He accepts AI performance only after comparing it to a human validated ground truth dataset and allows it to scale after reaching acceptable accuracy (e.g.,99\%).

\quotes{We do annotation for 50 papers. We do it manually. When we reach 99\% accuracy, and then we allow the system to do the other 1,000 papers}{--P8}

P3 mentioned that AI tools are very helpful in grant writing and paper writing.

\quotes{it has made things slightly easier in terms of like, organizing your thoughts, writing your papers, writing your grants, and other stuff. It helps you pretty well to restructure your grant in a way that would be useful for us.}{--P3}

\noindent In sum, the ways in which these tools, particularly genAI, are currently depend to a great extent on individual researchers' trust and adoption.





\begin{table*}[h]
\small
\caption{Perceptions, usage, and acceptance of AI and ML among social science researchers}
\begin{center}
\label{tab:subject1}
\begin{tabular}{|l|l|l|l|l|l|}
\hline
\textbf{Survey Question} &\textbf{Type}& \textbf{Mean} & \textbf{Sd} & \textbf{N} & \textbf{p-value} \\
\hline
\multirow{2}{*}{How familiar are you with the core concepts of AI/ML?} & ML & 2.72 & 1.02 & 141 & \multirow{2}{*}{\textbf{0.00012}} \\\cline{2-5}
 & AI & 3.19 & 1.01 & 143 &\\
\hline
\multirow{2}{*}{How frequently do you use AI/ML  during your research?} & ML & 2.09 & 1.13 & 141 & \multirow{2}{*}{\textbf{0.046}}\\\cline{2-5}
 & AI & 2.36 & 1.14 & 143 & \\
\hline
\multirow{2}{*}{To what extent do you accept the use of AI/ML in your research domain?} & ML & 3.49 & 1.02 & 141 &\multirow{2}{*}{\textbf{0.0001}} \\\cline{2-5}
 & AI & 2.98 & 1.16 & 143 &\\

\hline
\multirow{2}{*}{How useful do you think  AI/ML tools are for researchers in your field?} & ML & 3.37 & 0.85 & 141 &\multirow{2}{*}{\textbf{0.00014}}\\\cline{2-5}
 & AI & 2.95 & 0.98 & 143 & \\

\hline
\end{tabular}
\end{center}
\end{table*}

\section{RQ2: Ethical, methodological and practical concerns}
\subsection{Technological, operational and ethical concerns}
Both ML and AI raise ethical issues. However, in the ML survey, we observe that there are generally fewer ethical concerns. Most respondents to the AI survey expressed some level of concern or hesitation about the use of AI in social science, 
centering around potentially threatening foundational aspects of human-centered research. 

\quotes{I am okay with machine learning in which the research trains the models, but I am deeply skeptical of generative AI built on non-transparent and proprietary training data}{-- AI Survey}

\subsubsection{Limited understanding, training, and resources.}
In both surveys, several researchers acknowledged that they do not have sufficient knowledge about AI and ML to effectively evaluate their usefulness in social science research. This lack of knowledge extends not just to how these systems work technically but also to their appropriate application in research design, methodology, and interpretation. A dominant concern emerging from the responses is that individuals—particularly students and early-career researchers—may adopt these tools without a proper conceptual understanding of how these systems generate outputs. 

\quotes{Lack of understanding of relationship between ML and statistics, inadequate graduate training}{-- ML Survey}

In educational settings, this concern is amplified with the use of genAI. Students use AI to “cheat” rather than learn according to many respondents. Students who lean too heavily on AI tools might never learn to troubleshoot code, critically assess data, or synthesize information independently. Instead of gaining fluency in research methods, they risk becoming passive users of AI outputs. 

\quotes{My hesitation is students not understanding how to critically think for themselves.}{--AI survey}

\begin{table}[h]
\centering 
\small
\caption{Gender, rank, and field of interview participants}
\label{tab:subject1}
\begin{tabular}{|l|l|l|l|l|}
\hline
\multicolumn{1}{|c|}{\textbf{ID}} & \multicolumn{1}{|c|}{\textbf{Gender}} & \multicolumn{1}{|c|}{\textbf{Profession}} & \multicolumn{1}{|c|}{\textbf{Research field}} \\ \hline
P1  & Male & PhD student & Economics \\
P2  & Male   & PhD student  & Political Science \\
P3  & Female & PhD student & Psychology \\
P4  & Male & PhD student & Political Science\\
P5  & Female   & Associate Prof & Education \\
P6  & Male  &  PhD student  & Sociology  \\
P7  & Female & Assistant Prof &  Psychology \\
P8  & Male & PhD student  & Psychology \\
P9  & Female   & PhD student  & Psychology  \\
P10 & Male & Assistant Prof  & Sociology \\
P11 & Male & PhD student  & Economics   \\
P12 & Female   & PhD student  & Psychology  \\
P13 & Male & PhD student  & Economics  \\
P14 & Female   & Assistant Prof & Political Science        \\
P15 & Female & PhD Student & Sociology \\
\hline
\end{tabular}
\end{table}

According to one participant, these tools can artificially enhance the quality of a CV, which may not accurately reflect the candidate’s actual qualifications. 

\quotes{One of my students used ChatGPT to create her resume. She never reviewed the resume, so she just assumed that it was fine, and then asked me to write a recommendation letter for her, I found that the resume vastly oversold what she actually accomplished. She said she was writing papers, running her own analyses, and coming up with the theoretical models. Obviously, these things were not true. The embedded things here have huge ethical challenges for me. She trusts it so much that it's actually leading her to abandon critical thinking of the outputs intentionally.}{--P10}

\subsubsection{Ethical risks and bias.}
Many researchers expressed serious ethical concerns regarding the use of AI in social science research. Although ethical concerns are present in both responses they appeared more frequently in the AI survey. 

\quotes{The field is rapidly expanding and I worry about ethics informing that expansion. I also worry about bias and the diversity and quality of the training data.}{-- AI Survey}

Respondents highlighted that AI models are trained on existing data, which often reflects the historical and societal inequalities embedded in those datasets. For instance, they mentioned these tools used in facial recognition or predictive modeling have been shown to reproduce and even amplify racial, gender, and socioeconomic biases. This is particularly troubling in social science contexts, where research outcomes can influence policy decisions, social interventions, and public opinion. Several researchers pointed out that AI systems may inadvertently perpetuate discriminatory practices if applied uncritically, especially in sensitive areas like policing, surveillance, and healthcare.

\quotes{It's a black box that transmutes bias in the training data into bias in the outputs in a difficult to audit/detect way. It's also incredibly environmentally destructive.}{-- AI Survey}

Concerns were also raised about the ethical sourcing of training data. Some researchers questioned whether data used to train AI models were obtained with proper consent, particularly when derived from personal or publicly available information mentioned by P12. 

\quotes{“There are IRB concerns when it comes to what you would put into some of these models… run the risk of data sharing and privacy issues related to where that information is stored, how it is used, and how it is going to be used to train other models.”}{--P12}

Other growing concerns mentioned during the inverviews included: plagiarism; academic integrity; and misconduct in peer review. 
Interviewees noted ethical gray areas where it is unclear how to appropriately cite AI-generated content or how to ensure the originality of work. Some researchers worried this could lead to a dilution of scholarly standards, especially among students and less experienced scholars who might misuse AI-generated content without a clear understanding of academic norms. 

\quotes{We haven't reached the point of having clear norms about when LLMs are acceptable and not acceptable, including things like peer review. For example, I've spoken with colleagues and editors of journals who are getting, like AI-generated reviews. The use case that we just mutually agree that you know we should not use LLMs, is write a peer review}{--P14}

The cost of subscriptions are also seen as a barrier to equitable adoption: 

\quotes{Amongst people who use the tools, there's a possibility of inequity... there's a subscription model for the latest version of ChatGPT... \$40 a month, which is a lot of money.}{--P15}

Fabricating data, generating false citations, or creating papers that appear rigorous but lack substantive analysis could flood academic publication channels with low-quality or misleading research, undermining the credibility of scholarly communication. 

\quotes{Hype chasing is a major issue because it "sucks the oxygen" out of the room in terms of resources and time. Furthermore, it enables bad actors to plug up publication pipelines with garbage. Not to mention generative text and images are changing the incentive structures of producing content online -- it will be very difficult to find content actually produced by people who care about the content next to a mountain of botshit.}{-- AI Survey}

\subsubsection{Lack of standardization and best practice.}
A prominent concern voiced by researchers was the lack of standardization and established best practices in the use of AI tools for social science research. This issue was raised both explicitly and implicitly in multiple responses, reflecting a widespread unease about how AI is currently being integrated into academic workflows without clear guidelines or benchmarks. 
New tools, models, and capabilities are being released continuously, often without peer-reviewed validation or clarity around their appropriate application. This leaves researchers unsure about which tools are reliable, what constitutes appropriate use, and how to evaluate AI-generated outputs within the standards of rigorous social science inquiry. 

\quotes{Yes, I agree that there is currently a lack of standardization in terms of what can be used and what cannot be used}{-- AI Survey}

This also leads to concerns about transparency and documentation. Without clear standards, researchers may omit or inconsistently report how AI tools were used in their methods sections, making it difficult for peers to evaluate the integrity of the work or to reproduce the findings mentioned by 60\% of the interview participants as well. 




\subsubsection{Over-reliance and overuse.}
Over-reliance and overuse are major concerns reported across both surveys and interviews. Specific technologies such as AI-based decision support systems, autonomous systems, or recommendation systems show high rates of over-reliance compared to others. As users grew accustomed to the convenience and efficiency offered by automated tools, they started to begin to trust these systems excessively, often without verifying outputs or questioning decisions. This automation bias can reduce critical thinking skills, diminish situational awareness, and reduce the ability to detect errors, especially in complex or dynamic environments. 

\quotes{I think the overuse of ML has become more prevalent in research, even outside my field. As the ML technique is evolving so fast, some researchers might use it without understanding it, or use it when it is not necessary.--ML survey}

\subsubsection{Difficulty in model interpretability.}
Difficulty in model interpretability and high complexity are critical challenges highlighted in the data. As AI models, particularly deep learning systems, become more complex, they often operate as "black boxes," producing outputs without offering clear insights into how decisions are made. This lack of transparency can undermine user trust, hinder effective oversight, and complicate error diagnosis. 

\quotes{Intepretability is certainly key to much social scientific research, but ML models are increasingly complex and difficult to interpret. This is probably the biggest challenge.}{--ML Survey}


\subsubsection{Environmental harm.}
Environmental harm is a significant concern associated with the development and deployment of complex AI models, as reflected in the data. 
The intensive energy consumption required for model training and operation contributes to a large carbon footprint, exacerbating environmental degradation. 
As AI models grow larger and more complex, these costs—both monetary and ecological—continue to escalate, raising urgent ethical questions about sustainability.

\quotes{I have felt that there’s been a lack of acknowledgment of the environmental impact of using generative AI models. Not everyone needs to have an AI-generated image for their presentation.}{--P9}

\section{RQ3: Role of human judgment and interpretation in social science research}
We asked survey participants to envision scenarios where AI/ML complements or collaborates with human researchers in social science research. We followed up with a similar question during interviews. 

\subsection{AI/ML as collaborators}
Many respondents expressed that AI-powered tools are increasingly being viewed as valuable collaborators in social science research, particularly for tasks that involve handling large volumes of data or text. A majority indicated optimism about AI's potential to assist with data coding, statistical interpretation, literature summarization, and generating outlines or code skeletons for research projects. These tools were especially noted for their utility in navigating vast amounts of academic literature and supporting non-native English speakers with writing and proofreading. 

\quotes{Yes I think that its ability to expedite complicated, lengthy processes like literature reviews or brainstorming paper titles, etc. can allow researchers time to focus more on things like how to write in a way that allows for communication to a wide audience, not just scientific communities.}{--AI Survey}
    

\subsection{Human judgments are irreplaceable}
\subsubsection{Critical thinking and analytical reasoning.}
Participants consistently emphasized that critical thinking and analytical reasoning are fundamental human skills that remain irreplaceable in research. They highlighted that while AI and ML tools can process large amounts of data and identify patterns, they fundamentally (still) lack the capacity for deep reasoning, complex causal inference, and nuanced interpretation. Many noted that research demands not just pattern recognition, but the ability to critically assess evidence, weigh conflicting information and develop sophisticated theoretical insights capabilities that current AI systems cannot replicate. 


\quotes{Humans will be hardest to replace when normative judgments—about which questions are important and which policy solutions are appropriate—must be made.}{--AI Survey}

\subsubsection{Creativity, innovation, and theory development.}
While AI is seen as a potentially useful tool for tasks like measurement, data analysis, and hypothesis generation, participants believe that the core intellectual and theoretical contributions—such as theory building, interpreting political implications, and engaging in value discourse are domains where human insight cannot be replaced.


\quotes{Absolutely. Historical and social context are absolutely needed. Groundtruthing needs to be done by people. And theory development and writing can’t be automated, even if some ML tools might enhance our ability to do this work effectively and at scale.}{--ML Survey}

\subsubsection{Ethical reasoning and human-centered judgment.}
Participants consistently highlighted the irreplaceable role of human ethical reasoning and human-centered judgment in research. They noted that AI and machine learning systems currently lack the ability to navigate complex moral considerations, social norms, and the ethical dilemmas that often arise in research practice. Many stressed that ethical decision-making such as ensuring participant welfare, mitigating bias, interpreting sensitive contexts, and maintaining integrity requires human intuition, empathy, and moral responsibility. There was a strong sense that delegating ethical reasoning entirely to AI would risk undermining the foundational human values that guide responsible research. 

\quotes{Areas where human morality and direct interaction with other humans are needed. Unless AIs become extremely sophisticated, there will always be areas where takes humans to know humans.}{--AI Survey}

\subsubsection{Validation, verification, and quality control.}
Validation, verification, and quality control of research processes and outcomes have been viewed as the most important steps where human judgments are irreplaceable. Activities such as ground truth, critical review, and making final judgments were seen as areas where human oversight remains indispensable. Many participants expressed deep skepticism about the idea of fully trusting AI systems without human verification, pointing out that even highly advanced AI models can generate errors, overlook context, or reinforce biases. 

\quotes{I don't have a problem with AI. But we need the validation by humans. And I think if we are doing qualitative research, it becomes even more important to validate by humans.}{--P6}


\subsubsection{Motivation, passion, and human values.}
Motivation, passion, and human values are fundamental drivers of meaningful research, qualities that AI and ML cannot replicate. Several responses highlighted that passion fuels perseverance through complex, ambiguous, and challenging research problems, often inspiring creative breakthroughs that AI lacks the capacity to achieve. 

\quotes{Yes, human creativity, human passion, and judgment and interpretation are key.  AI isn't able to be creative and passionate. }{--AI Survey}


\section{Discussion and Recommendations}
AI and ML are reshaping research practices in the social sciences. While these technologies expand the technological toolkit available to researchers, they also simultaneously introduce new tensions around human judgment, automation, bias, and research integrity. The rapid adoption of generative AI tools, particularly large language models like ChatGPT, has enabled researchers to automate or accelerate tasks such as literature summarization, idea generation, coding support, and exploratory analysis. 
However, our findings show that this integration is not uniform in social science, while some researchers view genAI as a productivity tool, others remain cautious due to concerns about accuracy, deskilling, and ethical implications. ML techniques, on the other hand, are perceived as more statistically rigorous and better aligned with established research norms in fields that emphasize prediction and data modeling. Yet, they remain less accessible to researchers without technical training. This divergence reveals a dual dynamic: AI is seen as more usable but less trustworthy; whereas, ML is more trusted but less frequently used. Importantly, social scientists are not merely adopting these technologies, they are critically engaging with them, reflecting on how these tools fit within their disciplinary values of interpretation, transparency, and social accountability.

\subsection{Human-centered integration of AI/ML in social science research}

Social scientists’ ambivalence toward AI and ML reflects long-standing concerns about the agency of technical systems within knowledge production processes \cite{star1999ethnography, suchman2007human}. Rather than just functioning as neutral instruments, AI and ML models actively shape what counts as knowledge, whose perspectives are prioritized, and which research questions are considered tractable or valuable. 
As prior works on algorithmic mediation highlights \cite{gillespie2014relevance, sandvig2014auditing}, technical systems encode values, assumptions, and limitations that become naturalized through repeated use. Our results showcase that these concerns need to be explored and attended to, to ensure that tools' inherent weaknesses and biases are not overlooked. 

Our participants' worry about `theoretical misalignment' and `de-skilling' point to deep epistemic shifts: \textit{AI tools privilege data-driven, pattern-based reasoning may crowd out theory-driven, critical, and contextual approaches essential to social science inquiry}. This finding resonates with work in critical algorithm studies \cite{seaver2017algorithms, boyd2021data}, which argues that machine learning systems risk promoting empiricism without explanation, undermining traditional scholarly norms of causal reasoning and reflective critique.

Our results also demonstrate that while social scientists increasingly view AI/ML tools as valuable collaborators—especially for tasks involving data management, statistical analysis, and literature summarization—there is strong consensus that these technologies must remain subordinate to human judgment. This aligns with prior HCI work emphasizing human-in-the-loop approaches \cite{ehsan2021expanding}, which argue that AI systems should augment rather than automate human decision-making, especially in domains requiring interpretation, ethical reasoning, and creativity.
We highlight the persistent emphasis by participants on maintaining critical thinking, ethical oversight, and theoretical engagement. 

Thus, adopting AI in social science is not just a technical decision—it is a political and epistemological one that needs to be navigated carefully to preserve the disciplines' interpretive traditions. Stemming from the findings of STS theory, we argue that a potential way to move forward would rather be to focus on giving power back to the community by allowing organizations and technologies to be designed together, recognizing the interdependent relationship between technical and social systems \cite{walker2008review, narayanan2023towards}. The tenets of HCXAI (Human-centered Explainable AI) \cite{ehsan2021expanding, ehsan2020human, chakravorti2023prototype} can also be used to drive this initiative, putting the relevant community that interacts with the system at the center of technology design.

\subsection{Standardization, transparency, and accountability}
Another major theme identified through our work was the \textit{risk of automation bias}—an over-reliance on AI outputs at the expense of double-checking and validation. Participants' concerns reflect broader findings in human-AI interaction research, where trust calibration remains a core challenge \cite{dietvorst2015algorithm, zhang2020effect}. Over time, users tend to defer to technological and algorithmic recommendations even when they conflict with domain knowledge, particularly when systems are black boxes \cite{o2017weapons}.

In the context of academic research, automation bias threatens not only individual judgment but the collective development of knowledge. As noted by \cite{green2019disparate, bender2021dangers}, reliance on algorithmic outputs without contestation can institutionalize existing biases, marginalizing alternative epistemologies and reinforcing dominant narratives. Additional concerns around bias, environmental harm, and data exploitation surfaced throughout our study. Prior research has shown that these issues are often treated as externalities to be \textit{managed} after technical deployment \cite{selbst2019fairness, sambasivan2021everyone}. However, our participants' narratives confirm that ethical risks are infrastructural: they are embedded into how AI systems are built, trained, deployed, and normalized.

For instance, concerns about dataset opacity, algorithmic bias, and representational harms mirror findings from \cite{paullada2021data, dev2022measures, blodgett2022responsible}, who argue that data provenance and curation practices must be treated as first-order ethical issues. Similarly, worries about carbon costs echo the Green AI movement \cite{strubell2020energy, schwartz2020green}, emphasizing that technical efficiency should be weighed against environmental sustainability. 

We also argue that future AI tools for research need to be explicitly designed to foreground uncertainty, invite interrogation, and support counterfactual exploration—design principles that align with calls for contestable AI \cite{binns2018fairness, alfrink2023contestable, reckinger2018situative}. Embedding friction into AI systems—through transparency prompts, interpretability features, or required human-in-the-loop validation—can help sustain critical rather than erode it. 

Finally, the absence of best practices for documenting and evaluating AI/ML-assisted research emerged as another barrier to trust among social science researchers. This adds to the concerns about the reproducibility crisis in computational fields \cite{sandve2013ten, chakravorti2025reproducibility, wu2024integrating} and points to the need for standardized methodologies. Drawing from frameworks proposed in explainable AI (XAI) \cite{miller2019explanation} and HCI for ML systems \cite{amershi2019guidelines}, we argue that social science research domains need to develop AI documentation protocols that include: 
[1] model transparency (e.g., architecture, training data, limitations); [2] usage disclosure (e.g., which parts of the analysis involved AI assistance); [3] evaluation criteria (e.g., audits for fairness, bias, epistemic validity); and [4] reproducibility artifacts (e.g., code, data, prompt logs). 

\subsection{Recommendations: A way towards responsible integration of AI in the social sciences} 

Taking inspiration from the findings provided by our participants, as well as motivated by prior technological audits and recommendation frameworks \cite{blodgett2022responsible, bender2018data, venkit2024confidently}, we provide \textit{research-driven recommendations} for different stakeholder groups in this social science space involved in the integration of AI and ML into social science research ecosystems.

\paragraph{For AI Developers and Tool Designers}
\begin{itemize}
    \item \textbf{Prioritize Explainability and Transparency:} Incorporate interpretable models, visual explanations, and clear documentation of training data and model behavior.
    \item \textbf{Incorporate Lived Experiences into the Design Process:} Involve end users, especially social science researchers—in ongoing, real-world testing and iterative feedback loops to ensure AI tools reflect actual needs, values, and research practices. 
    \item \textbf{Enable Critical Interventions:} Provide users with editable outputs, override mechanisms, and configurable prompt logs for agency and accountability.
    \item \textbf{Embed Ethical Safeguards:} Integrate prompts, alerts, or design constraints that flag potentially biased, unsafe, or misleading outputs, if possible.
    \item \textbf{Support Sustainability:} Disclose environmental impacts (e.g., carbon emissions) and develop lightweight alternatives where possible.
\end{itemize}

\paragraph{For Researchers and Social Science Practitioners}
\begin{itemize}
    \item \textbf{Develop Critical AI Literacy:} Engage with foundational technical knowledge and ethical frameworks to understand the capabilities and limitations of AI/ML systems.
    \item \textbf{Adopt Human-AI Collaboration Mindsets:} Use AI as an assistant rather than a substitute; maintain human judgment in interpretation, validation, and theorizing.
    \item \textbf{Engage in Participatory, Lived Experimentation:} Treat the implementation of AI/ML tools not as static adoption but as dynamic, co-constructed processes, using lived experiments to surface tacit knowledge, contextual concerns, and ethical challenges.
    \item \textbf{Document AI Usage Transparently:} Clearly describe when, where, and how AI tools are used in research to enhance reproducibility and trust.
    \item \textbf{Maintain Ethical Oversight:} Continuously assess representational harms, dataset bias, and implications for marginalized populations in research contexts.
\end{itemize}

\paragraph{For Educators and Academic Institutions}
\begin{itemize}
    \item \textbf{Integrate Critical AI Pedagogy:} Incorporate algorithmic ethics, bias analysis, and sociotechnical systems thinking into curricula for students in both technical and non-technical disciplines.
    \item \textbf{Support Interdisciplinary Skill-Building:} Offer workshops, boot camps, and joint seminars that bring together social scientists, data scientists, and designers.
    \item \textbf{Revise Research Training Guidelines:} Update review board (eg: IRB) protocols and academic writing guidelines to include AI-specific considerations.
\end{itemize}

\paragraph{For Policymakers and Funding Bodies}
\begin{itemize}
    \item \textbf{Support Development of Ethical Standards:} Fund the creation of domain-specific AI governance guidelines, audit frameworks, and participatory design protocols.
    \item \textbf{Mandate Sustainability Reporting:} Require AI research proposals to include environmental impact assessments and encourage the adoption of carbon-conscious modeling practices.
    \item \textbf{Promote Equitable Access:} Ensure equitable distribution of AI infrastructure, funding, and training opportunities across institutions and researcher demographics.
    \item \textbf{Encourage Community-Led Governance:} Involve diverse stakeholders, including marginalized communities, in decisions about AI deployment and oversight in research. 
\end{itemize}

\section{Conclusion}
Our work highlights that ways in which the integration of AI and ML into social science research marks a significant sociotechnical shift raising important ethical and methodological questions. Through a combination of surveys followed by semi-structured interviews, we found that AI especially in its generative AI form is increasingly accessible and attractive to everyone. Its rapid adoption raises serious concerns. These include over-reliance, erosion of theoretical grounding, risks of bias, threats to research integrity, transparency, and environmental impacts. Meanwhile, machine learning, although more rigorously aligned with statistical reasoning, remains less accessible to practitioners due to complex methodologies. This divergence in accessibility and perception underscores the need for targeted capacity building, clearer standards, and ongoing ethical reflection. Critically, we have argued for the importance of human-centered designs--not just a matter of interface design but a deeper commitment to preserving the agency, expertise, and interpretive traditions of social science. Ethical engagement, methodological transparency, and inclusive design must be core to how AI and ML systems are built and adopted. This means moving beyond technological determinism toward a more participatory, reflective, and justice-oriented framework of research practice. In response, we offer actionable recommendations for developers, researchers, educators, and policymakers focused on explainability, ethics, sustainability, and equitable access. 

\section{Ethical Considerations Statement}
Our work adheres to ethical standards. Survey and interview design and the data collection were approved by the [anonymized] University IRB department before the recruitment of the human subjects. Participants were fully informed about the nature of the study, potential risks, and their right to withdraw at any time without penalty before the study began. All participants provided informed consent prior to engaging with the survey and interview, and their data were anonymized to protect privacy. We have only considered responses where the participants have given consent to use and analyze the data. 

\section{Positionality Statement}
The research team, with diverse expertise in social sciences and engineering disciplines including Human-Computer Interaction, Computer Engineering, Political Science, and Education Research, contributed to a multifaceted analysis of the data. All authors actively contributed to the interpretation of the findings and discussions regarding the implications of the study. Our perspectives are informed by our experiences in both academia and industry, particularly in the development of human-centered AI systems and providing recommendations. To mitigate personal biases and avoid over-interpreting the data, we documented and critically examined our preconceptions throughout the study. This reflexivity ensured a more balanced approach to analyzing and interpreting our findings.

\bibliography{aaai24}

\end{document}